\documentclass[manuscript]{emulateapj}

\shorttitle{H.E.S.S. observations of GRB~060602B}
\shortauthors{Aharonian et al.}

\begin{document}

\title{H.E.S.S. observations of the Prompt and Afterglow
  Phases of GRB~060602B}
\author{F. Aharonian,\altaffilmark{2,14}
 A.G.~Akhperjanian,\altaffilmark{3}
 U.~Barres de Almeida,\altaffilmark{9,30}
 A.R.~Bazer-Bachi,\altaffilmark{4}
 B.~Behera,\altaffilmark{15}
 M.~Beilicke,\altaffilmark{5}
 W.~Benbow,\altaffilmark{2}
 K.~Bernl\"ohr,\altaffilmark{2,6}
 C.~Boisson,\altaffilmark{7}
 V.~Borrel,\altaffilmark{4}
% ...
}

% ...
\altaffiltext{2}{Max-Planck-Institut f\"ur Kernphysik, Heidelberg, Germany}
\altaffiltext{3}{Yerevan Physics Institute, Armenia}
\altaffiltext{4}{Centre d'Etude Spatiale des Rayonnements, CNRS/UPS, Toulouse, France}
\altaffiltext{5}{Universit\"at Hamburg, Institut f\"ur Experimentalphysik, Germany}
\altaffiltext{6}{Institut f\"ur Physik, Humboldt-Universit\"at zu Berlin, Germany}
\altaffiltext{7}{LUTH, Observatoire de Paris, CNRS, Universit\'e Paris Diderot, France}
% ...
\altaffiltext{9}{University of Durham, Department of Physics, UK}
% ...
\altaffiltext{14}{Dublin Institute for Advanced Studies, Ireland}
\altaffiltext{15}{Landessternwarte, Universit\"at Heidelberg, K\"onigstuhl, Germany}
% ...
\altaffiltext{30}{supported by CAPES Foundation, Ministry of Education of Brazil}

\begin{abstract}
We report on the first completely simultaneous observation of a gamma-ray burst (GRB) using an array of Imaging Atmospheric Cherenkov Telescopes which is sensitive to photons in the very-high-energy (VHE) $\gamma$-ray range ($\gtrsim100$~GeV). On 2006 June 2, the \emph{Swift} Burst Alert Telescope (BAT) registered an unusually soft $\gamma$-ray burst (GRB~060602B). The burst position was under observation using the High Energy Stereoscopic System (H.E.S.S.) at the time the burst occurred. Data were taken before, during, and after the burst. A total of 5 hours of observations were obtained during the night of 2006 June 2--3, and 5 additional hours were obtained over the next
3 nights. No VHE $\gamma$-ray signal was found during the period covered by the H.E.S.S. observations. The 99\% confidence level flux upper limit ($>$1~TeV) for the \emph{prompt} phase (9~s) of GRB~060602B is $2.9\times10^{-9}\,\mathrm{erg\,cm^{-2}\,s^{-1}}$. Due to the very soft BAT spectrum of the burst compared to other \emph{Swift} GRBs and its proximity to the
Galactic center, the burst is likely associated with a Galactic
X-ray burster, although the possibility of it being a cosmological GRB cannot be ruled out. We
discuss the implications of our flux limits in the context of these two bursting scenarios.
\end{abstract}


\section{Introduction}

    Gamma-ray bursts (GRBs) are brief and intense flares of $\gamma$-rays. Without precedent in astronomy, they arrive from random directions in the sky and last typically~$\sim$0.1--100~s~\citep[\emph{prompt} emission, see][]{klebesadel73,fishman95}. The very nature of GRBs makes it operationally rather challenging to study their \emph{prompt} phase simultaneously in any other wavelength.

    The observed GRB properties are generally well explained by the \emph{fireball} model, in which the emission is produced in relativistic shocks~\citep{piran99,zhang04,meszaros06}. In this standard model, the highly-relativistic plasma, which emits the observed sub-MeV radiation, is expected to generate $\gamma$-rays up to the very-high-energy (VHE; $\gtrsim$100~GeV) regime, via inverse-Compton emission of electrons or proton-induced mechanisms~\citep{zhang01,peer05,asano07,fan08}. Therefore, the detection of gamma-rays or
    sufficiently sensitive upper limits would shed light on our understanding of the current model. Some important yet largely unknown parameters in GRB models, such as the bulk Lorentz factor and the opacity of the outflow just after the acceleration phase, can be directly measured through high-energy (HE; $\ga$100 MeV) and VHE $\gamma$-ray observations during the \emph{prompt} phase of GRBs~\citep{razzaque04,baring06}.

    There are two techniques used in VHE $\gamma$-ray astronomy to observe the \emph{prompt} phase: the first technique is to slew quickly to the GRB position provided by a burst alert from
    satellites. This technique is used for Imaging Atmospheric Cherenkov Telescopes (IACTs), such as the High Energy Stereoscopic System (H.E.S.S.), which have a field of view (FoV) of a few degrees. The MAGIC telescope, operating in this mode, was able to slew to the position
    of GRB~050713A, 40~s after the GRB onset, while the \emph{prompt} keV emission was still active. A total of 37 minutes of observations were made and no evidence of emission above 175~GeV was obtained~\citep{albert06a}. The rapid follow-up observations using this telescope of 8 other GRBs show no evidence of VHE $\gamma$-ray emission from these GRBs during the \emph{prompt} or the \emph{early afterglow} phase~\citep{albert07}. However, there is always a delay in time for IACTs operating in this GRB-follow-up
    mode, as long as the GRB position lies outside the camera FoV at the onset of the GRB. This results in an incomplete coverage of the GRB \emph{prompt} phase.

     The second technique is to observe a large part of the sky continuously, at the expense of much lower sensitivity than the IACT detectors. This technique is used, e.g. for the water Cherenkov detector Milagro, which works at higher energies than current IACTs. Since the effect of extra-galactic background light (EBL) absorption increases with the energy of a $\gamma$-ray photon, the higher energy threshold of Milagro thus lowers its chance to detect VHE $\gamma$-rays from distant GRBs, when compared to IACT detectors. No evidence of VHE $\gamma$-ray emission was seen from 39 GRBs using this detector~\citep{atkins05,abdo07}. \citet{atkins00} reported a possible VHE $\gamma$-ray enhancement coincident with GRB~970417A (with a post-trials probability $1.5\times10^{-3}$ of being a background fluctuation) using Milagrito, the forerunner of Milagro.

    In this paper, we report on the first completely simultaneous observation with an IACT
    instrument of a $\gamma$-ray burst (GRB~060602B) using H.E.S.S. The burst position fell serendipitously at the edge of the FoV of the H.E.S.S. cameras when the burst occurred.

\section{GRB 060602B}

\begin{figure}
    \epsscale{.9}
   \plotone{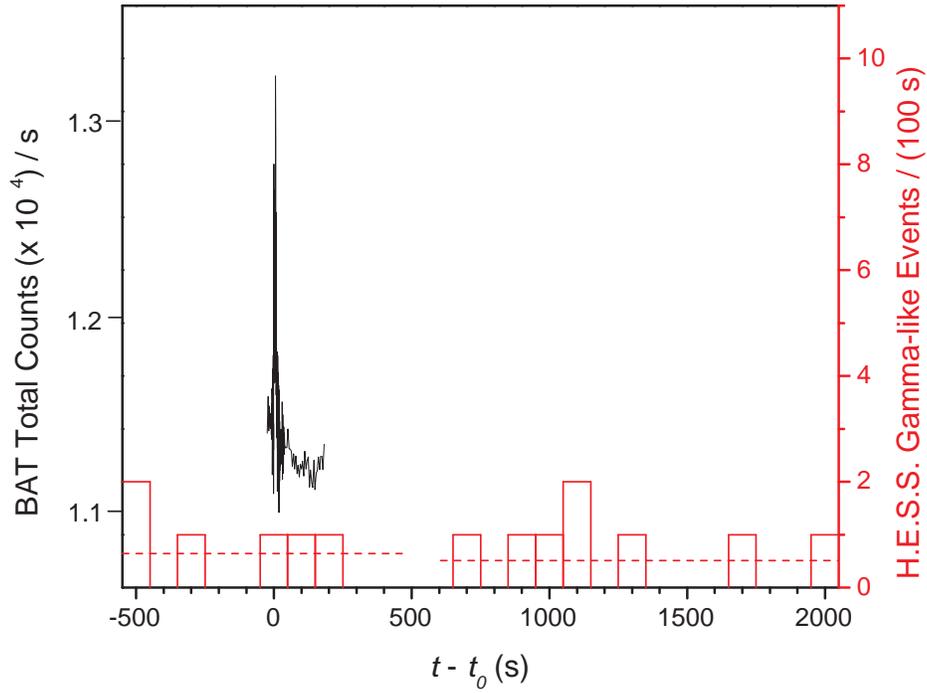}
      \caption{\emph{Histograms and right scale}: Gamma-like events, i.e. those that passed standard cuts, as observed using H.E.S.S. within a circular region of radius $\theta_{\rm cut}=0.32\arcdeg$ (for $t < t_0+500$s, with a large offset, see text) and $\theta_{\rm cut}=0.11\arcdeg$ (for $t > t_0+600$s) centered at the burst position. The dashed horizontal lines indicate the expected number of background events in the circular regions, using the reflected-region background model~\citep{berge07}. The gap between $\sim$500s and 600s is due to a transition between observation runs. \emph{Solid curve and left scale}: \emph{Swift}/BAT light curve in the 15-150 keV band.}
         \label{simultaneous}
   \end{figure}

At 23:54:33.9 UT on 2006 June 2 (denoted by $t_0$), the Burst Alert Telescope (BAT) on board \emph{Swift}, which operates in the $15-350$ keV energy band, triggered on GRB~060602B~\citep[trigger 213190,][]{schady5200}. The refined BAT position was R.A.~$=17\mathrm{^h}49\mathrm{^m}28.2\mathrm{^s}$, Dec.~$=-28\arcdeg7\arcmin15.5\arcsec$ \citep[J2000;][]{palmer5208}. The BAT light curve showed a single-peaked structure lasting from $t_0-1\,\mathrm{s}$ to $t_0+9\,\mathrm{s}$ (Figure~\ref{simultaneous}). The peak was strongest in the 15--25~keV energy band and was not detected above 50~keV. $T_{90}$ (defined as the time interval between the instants at which 5\% and 95\% of the total integral emission is detected in the 15--350 keV band) was $9 \pm 2\,\mathrm{s}$~\citep{palmer5208}. This $\sim$9-s time interval is referred to as the \emph{prompt} phase of this GRB in this work. \citet{palmer5208} fit the time-averaged energy spectrum from $t_0-1.1\,\mathrm{s}$ to $t_0+8.8\,\mathrm{s}$ by a simple power law with a photon index of $5.0 \pm 0.52$, placing it among the softest of the \emph{Swift} GRBs. Using the data from the same time interval, a 15--150 keV fluence of $(1.8 \pm 0.2) \times 10^{-7}\,\mathrm{erg\,cm}^{-2}$ was derived. No spectral evolution was observed during the burst~\citep{wijnands08}.

\emph{Swift}'s other instrument, the X-ray Telescope (XRT), began data-taking 83~s after the BAT trigger and found a fading source. \citet{beardmore5209} reported a position R.A.~$=17\mathrm{^h}49\mathrm{^m}31.6\mathrm{^s}$, Dec.~$=-28\arcdeg8\arcmin3.2\arcsec$ (J2000), confirmed by later analyses~\citep{butler07,wijnands08}. This position (with an error circle of radius~$\sim$3.7$\arcsec$) was used in analyses presented in this paper. The flux faded temporally as a power law with an index of $0.99 \pm 0.05$ from $\sim t_0+100\,\mathrm{s}$ up to $\sim t_0+10^6\,\mathrm{s}$~\citep{wijnands08}.

\begin{figure}
    \epsscale{1.}
   \plotone{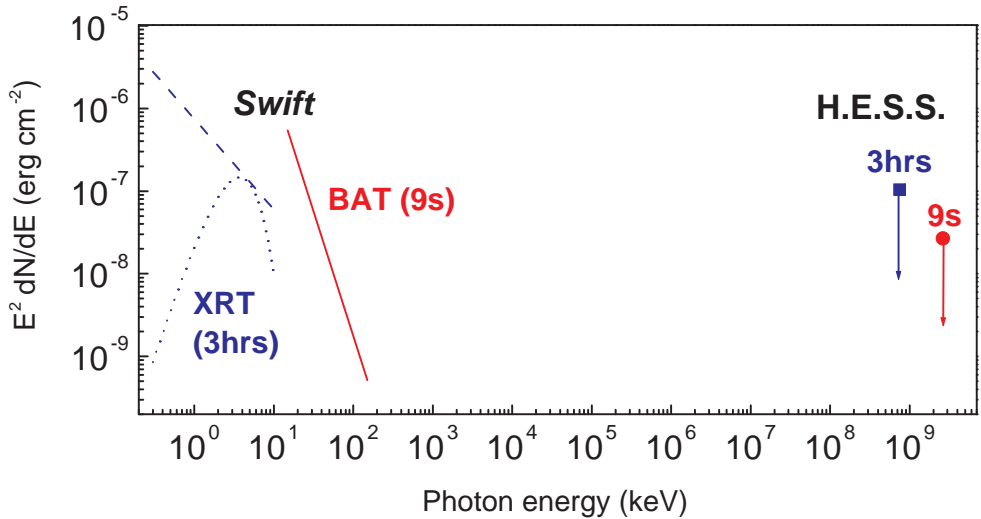}
      \caption{Time-integrated spectral energy distributions at the burst position during the 9-s \emph{prompt} phase and during the 3-hour \emph{afterglow} phase. A power-law model fitted to the BAT spectrum during
      the 9-s burst (solid line) is shown, as well as the \emph{source} spectra used in an absorbed power-law model (dashed line) and an absorbed blackbody model (dotted line) to describe the XRT spectrum during $100\,\mathrm{s}-11.4\,\mathrm{ks}$ after the burst onset. The H.E.S.S. upper limits derived from 9-s \emph{prompt} data (circle) and 3-hour \emph{afterglow} data (square) are also indicated. The H.E.S.S. \emph{prompt} and \emph{afterglow} limits are plotted at the corresponding average photon energies.}
         \label{sed}
   \end{figure}

Using data taken from $t_0+100\,\mathrm{s}$ to $t_0+11.4\,\mathrm{ks}$, the time-averaged 0.3--10~keV energy spectrum was fitted by an absorbed power-law model, $dN/dE\propto E^{-\Gamma_\mathrm{X}}$, where $E$ is the photon energy in keV and $\Gamma_\mathrm{X}$ the photon
index. The fit results in $\Gamma_\mathrm{X}=3.1^{+0.7}_{-0.6}$ and an absorption column density of $N_\mathrm{H}=4.6^{+1.6}_{-1.4}\times 10^{22} \mathrm{cm}^{-2}$, with $\chi^2/\mathrm{d.o.f}=34/35$. Fitting the same spectrum with an absorbed blackbody model, $dN/dE\propto E^2/[(kT)^4(e^{E/kT}-1)]$, a temperature of $kT=0.94^{+0.15}_{-0.13}$~keV and $N_\mathrm{H}=1.5^{+1.0}_{-0.9} \times 10^{22} \mathrm{cm}^{-2}$ were obtained, with $\chi^2/\mathrm{d.o.f}=36/35$. These two modeled \emph{source} spectra are shown in Figure~\ref{sed}, for comparison with the H.E.S.S. upper limits obtained over a comparable time interval. While the modeled
\emph{source} spectra look very different, after
different levels of absorption along the line of sight, they both
describe the observed data equally well, as shown by the normalized
$\chi^2$ values both close to 1. These results are consistent with the analyses of other authors~\citep{beardmore5209,wijnands08}.

In the optical or infrared band, no counterpart was found by the observations of several telescopes~\citep{kubanek5199,khamitov5205,blustin5207,melandri5229}. This is expected because of the severe optical extinction along this line of sight.

\section{The H.E.S.S. Observations}

The H.E.S.S. array is a system of four 13m-diameter IACTs located in the Khomas Highland of Namibia~\citep{hinton04}. The system has a point source sensitivity above 100~GeV of $\sim4\times10^{-12}\,\mathrm{erg}\,\mathrm{cm}^{-2}\, \mathrm{s}^{-1}$ (about $1\%$ of the flux from the Crab nebula) for a $5 \sigma$ detection in a 25~hour observation. The cameras of the H.E.S.S. telescopes
detect Cherenkov photons over a 5$\degr$ FoV, thus enhancing its ability to detect serendipitous sources, as demonstrated in the Galactic plane survey~\citep{aha05a}.

The position of GRB~060602B was under observation using H.E.S.S. before
the burst, throughout the duration of the burst, and after the
burst. The observations are shown in
Table~\ref{obspattern}. The zenith angles (Z.A.) and the offsets of the GRB~060602B position
from the center of the FoV are shown for each observation period.
A total of 4.9 hours of observations were obtained during the night of
2006 June 2--3. This includes 1.7~hour \emph{pre-burst}, 9~s \emph{prompt}, and 3.2~hour \emph{afterglow} phases. Additionally, 4.7 hours of observations at the burst position were
obtained over the next 3 nights. All data were taken in good
weather conditions and with good hardware status. The observations were taken with the GRB~060602B position placed at different offsets relative to the center of the FoV of the telescopes, because most observations were not dedicated to the position of GRB~060602B. The position offsets were rather large ($\geq 2.5\arcdeg$) during the period before the burst until $\sim$9~minutes after the burst.

Due to the H.E.S.S. long term monitoring program of the Galactic center region, a deep exposure of the GRB~060602B position (over a period of several years) also exists (see \S \ref{sect_results}).


\section{H.E.S.S. Data Analysis}

\begin{figure}
    \epsscale{.8}
    \plotone{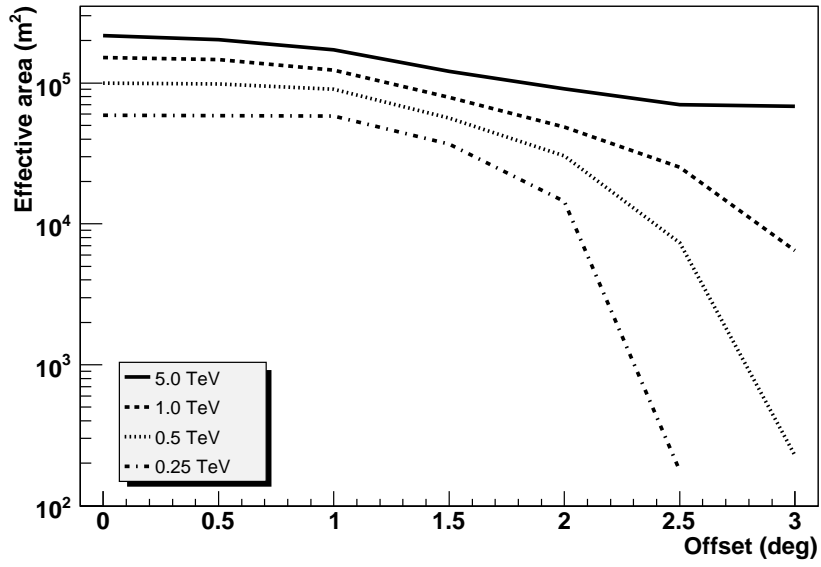}
      \caption{The effective areas for various photon energies at offsets from 0$\degr$ to 3$\degr$ from the center of the FoV for Z.A.$=0\degr$, using the standard cut analysis used in this work}
         \label{eff_area}
   \end{figure}

    Calibration of data, event reconstruction and rejection of the cosmic-ray background (i.e. $\gamma$-ray event selection criteria) were performed as described in~\citet{aha06a}, which employ the techniques described by~\citet{hillas96}. Targets are typically observed at a {\em normal}
    offset from the FoV center of 0.5$\degr$ or 0.7$\degr$ (\emph{wobble} mode), to allow for a simultaneous
    background estimate from regions in the FoV that have identical properties
    as the source position. At \emph{normal} offsets, the point spread function (PSF) and effective area for $\gamma$-rays are nearly identical to the values at the FoV center, according to air-shower simulations. However, the reconstructed event directions are less accurate at larger offsets. The PSF at the maximum offset of 2.9$\degr$ is by a factor of $\sim$2
    more extended than the one at \emph{normal} offsets. Figure~\ref{eff_area} shows the effective areas for various photon energies at offsets from 0$\degr$ to 3$\degr$ from the center of the FoV for Z.A.$=0\degr$, using the standard cut analysis described below.

    Gamma-like events were then taken from a circular region of radius $\theta_{\rm cut}$ centered at the burst position. The background was estimated using the reflected-region background model as described in~\citet{berge07}.

    Two sets of analysis cuts were applied to search for a VHE $\gamma$-ray signal. These include standard cuts~\citep{aha06a} and soft cuts (with lower energy thresholds, as described in~\citet{aha06b}\footnotemark). Standard cuts are optimized for a source with a photon index of $\Gamma=2.6$. Soft cuts are optimized for sources with steep spectra ($\Gamma=5.0$), thus having a better sensitivity at lower energies. The latter is useful for a source at cosmological distances, since the EBL absorption would greatly soften the intrinsic spectrum of the VHE $\gamma$-ray radiation from the source. For observational periods with a position offset of $2.9\arcdeg$, a larger $\theta_{\rm cut}$ value of $0.32\arcdeg$ was used to accommodate the larger PSF. Energy thresholds ($E_{\rm th}$) obtained for a standard cut analysis in each period are shown in Table~\ref{obspattern}.

    \footnotetext{\emph{Soft cuts} were called \emph{spectrum cuts} in~\citet{aha06b}.}

    Figure~\ref{simultaneous} shows the rate of $\gamma$-like events (i.e. those that passed standard cuts) observed within a circular region of radius $\theta_{\rm cut}=0.32\arcdeg$ (for $t < t_0+500$s) and $\theta_{\rm cut}=0.11\arcdeg$ (for $t > t_0+600$s) centered at the source.

     The independent \emph{Model} analysis technique~\citep{deNaurois05} was used to analyze the same data. The results obtained from both analyses are consistent with each other. Hence, only the analysis results based on Hillas parameters are presented in this paper.

\section{Results}
\label{sect_results}
\begin{figure}
    \epsscale{1.}
    \plotone{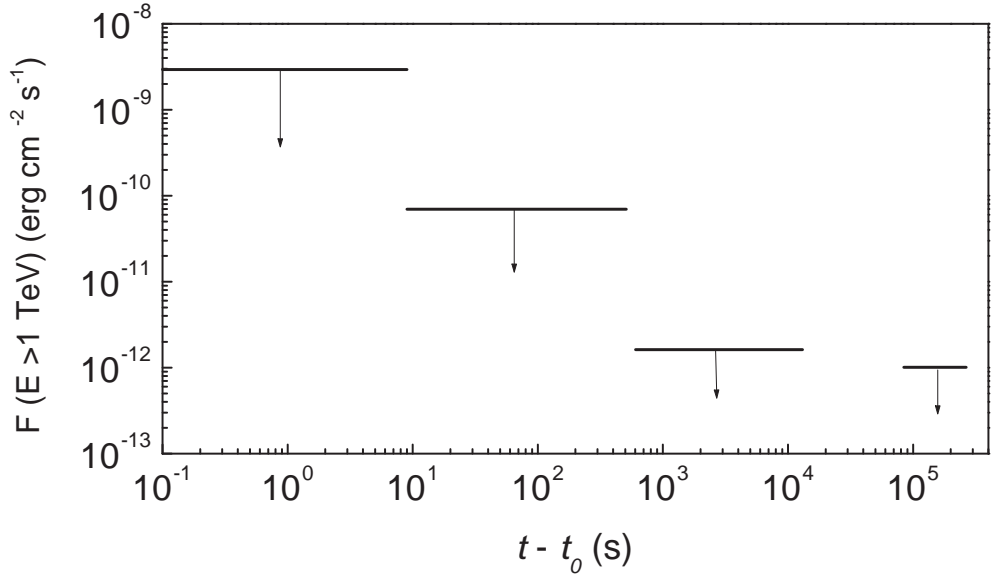}
      \caption{The 99\% confidence level flux upper limits at energies $>1\,\mathrm{TeV}$ derived from H.E.S.S. observations at the position of GRB~060602B during the \emph{prompt} and \emph{afterglow} phases. The two ends of the horizontal lines indicate the start time and the end time of the observations from which the upper limits were derived.}
         \label{lightcurve}
   \end{figure}

No evidence for excess $\gamma$-ray events was found at any time
before, during, or after the event GRB~060602B. A \emph{Crab-like} photon spectral index of 2.6 is assumed when deriving the flux limits presented in this section. The 99\% confidence level flux upper limits obtained by the method of~\citet{feldman98}
for every observation run using standard cuts are included in
Table~\ref{obspattern}. Figure~\ref{lightcurve} shows the 99\% energy flux upper limits above $1$~TeV during the \emph{prompt} and \emph{afterglow} phases up to 4 nights after the burst. The energy flux limit ($>$1~TeV) for the \emph{prompt} phase of GRB~060602B is $2.9\times10^{-9}\,\mathrm{erg\,cm^{-2}\,s^{-1}}$. The limits for the period $\sim10^2-10^4 \mathrm{s}$ after the burst are at levels comparable to the X-ray energy flux as observed by \emph{Swift}/XRT during the same period. These limits are not very sensitive to the assumed photon spectral index (within a factor of 2 when changing the index to 2 or 4).

H.E.S.S. observations from 2004 to 2006 covering the position of GRB~060602B
are used to constrain the time averaged emission from this object. No signal was found in the 128 hours of available data, of which more than 80\% were taken before the burst. Assuming constant emission,
a 99\% flux upper limit (using standard cuts) of $9.0\times
10^{-13}\,\mathrm{erg\,cm^{-2}\,s^{-1}}$ above 200 GeV (about 0.5\%
of the Crab flux) was found. This result is relevant for the Galactic scenario discussed in \S \ref{sect_XRB_scenario}.

Figure~\ref{sed} shows the spectral energy distribution of the
burst during the first 9~s, and during the period $t_{0} + 100\,\mathrm{s}$ to $11.4\,\mathrm{ks}$ ($\sim$3~hours) after the burst onset. It can be seen that the VHE energy fluence limits are of the similar level as the fluence at keV energies measured by \emph{Swift} for both the 9-s \emph{prompt} and 3-hour \emph{afterglow} phases. Due to the soft keV spectra, any radiation in the VHE range would very likely come from a high-energy component separated from that of the sub-MeV radiation.



\clearpage
\section{Discussion}

The nature of GRB 060602B is unclear. The softness of the BAT spectrum and the proximity of GRB~060602B to the Galactic center suggest a possible Galactic origin of the event. The observed temperature of $\sim$1~keV (using an absorbed blackbody fit) using XRT data is within the range seen from type-I X-ray bursts~\citep{kuulkers03}. The \emph{Swift}/BAT team has consequently classified the event as an X-ray burst~\citep{barthelmy6013}. \citet{halpern5210} noted that a faint source had been visible in an {\it XMM-Newton} observation taken in the neighborhood of the GRB~060602B position. Two other {\it XMM-Newton} observations were performed almost four months after the burst and a faint source was detected. The position of the faint source is marginally consistent with the {\it Swift}/XRT position of GRB~060602B, within the large positional errors~\citep[up to $4\arcsec$,][]{wijnands08}. However, no indication of variability of the source was seen and no secure spatial association of the source with GRB~060602B was established. 

Although a Galactic origin is more likely, the possibility of the GRB as a cosmological GRB is not ruled out. In this section, we briefly discuss the implications of the H.E.S.S. observations according to these two scenarios.

\subsection{Implications for the cosmological gamma-ray burst scenario}
\label{sect_GRB_scenario}
HE $\gamma$-ray emission have been detected in the \emph{prompt} and/or \emph{afterglow} phases of several GRBs~\citep{hurley94,gonzalez03,kaneko08}. In these cases, no evidence for a high-energy cut-off was seen. The temporal evolution of the HE emission of GRB~941017 was found to be significantly different from its low-energy $\gamma$-ray light curve~\citep{gonzalez03}. For GRB~970417A, if the excess events observed by Milagrito were actually associated with the burst, the photon energy must be at least 650 GeV and the VHE $\gamma$-ray energy fluence must be at least an order of magnitude higher than the 50-300 keV energy fluence as seen by BATSE~\citep{atkins03}.

In the VHE regime, possible radiation mechanisms include leptonic scenarios: external-shock accelerated electrons up-scattering self-emitted photons~\citep{dermer00,zhang01} or photons from other shocked regions~\citep{wang01,wang06}, and hadronic scenarios: proton synchrotron emission~\citep{boettcher98,totani98a,totani98b} or cascades initiated by $\pi^0$ produced via photo-meson interactions~\citep{boettcher98,waxman00}. In leptonic models, one typically expects a positive correlation between X-ray flux and VHE $\gamma$-ray flux. We note that the X-ray emission as seen by XRT decayed quickly, so one might expect the strongest VHE $\gamma$-ray emission to occur during the \emph{prompt} phase or soon after. In fact, during the \emph{early afterglow} phase, some authors predict VHE $\gamma$-ray energy flux levels comparable to or even higher than those in X-rays~\citep{wang01,peer05}.

The energy threshold of the H.E.S.S. observations was about 1~TeV and 250~GeV during the \emph{prompt} and \emph{afterglow} phases, respectively. For a cosmological GRB, VHE $\gamma$-ray radiation is attenuated by the EBL. The optical depth, $\tau$, of the EBL absorption for a 1~TeV and 250~GeV photon is about unity at $z=0.1$ and $0.3$, respectively~\citep{aha06d}. Therefore, if GRB~060602B occurred at $z\la0.2$, EBL absorption could be neglected. Under this assumption, the H.E.S.S. flux limits would exclude an intrinsic VHE $\gamma$-ray \emph{prompt} and \emph{afterglow} energy fluence much higher than that at sub-MeV energies (see Figure~\ref{sed}). Also, a VHE $\gamma$-ray fluence level such as the one implied by the possible $\gamma$-ray events associated with GRB~970417A would be excluded for GRB~060602B. And the upper limits would constrain models which predict VHE $\gamma$-ray energy flux levels higher than those in X-rays during $\sim10^2-10^4 \mathrm{s}$ after the burst. If, however, GRB~060602B occurred at $z\ga0.2$, EBL absorption would be more severe and the observed limits would have to be increased by a factor which depends both on the redshift and the detailed gamma-ray spectrum of the GRB. In this case, the limits would be less constraining.

\subsection{Implications for the Galactic X-ray binary scenario}
\label{sect_XRB_scenario}
X-ray binaries have been suspected to be VHE $\gamma$-ray emitters for decades,~see, e.g. the review by~\citet{weekes92}, and have recently been confirmed for at least three cases~\citep{aha05b, aha06c, albert06b}.

Type-I X-ray bursts, originating from low-mass X-ray binaries (LMXBs) and with typical duration of 10~s up to several minutes, are caused by thermonuclear flashes on the surface of accreting neutron stars\footnote{This process was proposed to explain the origin of GRBs~\citep[see, e.g.][]{Hameury82,Woosley82}.}~\citep{lewin93}. Although most X-ray bursts are detected from known X-ray sources or transients, some X-ray bursts originated from the so-called \emph{burst-only} sources, whose quiescent X-ray luminosity is too low to be detected by current X-ray detectors~\citep{corn04}.

Based on the BAT spectrum of the burst and the possible identification of a faint \emph{XMM-Newton} X-ray counterpart, \citet{wijnands08} prefer the type-I X-ray burst scenario. In this case, the source might have been active in X-rays before the BAT trigger, although there was no detection with the {\it RXTE}/ASM before the burst~\citep{wijnands08}. The GRB~060602B position had been in the FoV of H.E.S.S. for $\sim$2~hours when BAT triggered the event. No significant VHE $\gamma$-ray emission was observed during this period. If this scenario is true, the H.E.S.S. observations rule out that this X-ray burst was accompanied by a VHE $\gamma$-ray burst of similar energy flux. To our knowledge, no simultaneous VHE $\gamma$-ray observation of a type-I X-ray burst has been reported. \citet{aha98} reported a tentative evidence of a possible TeV burst emission with HEGRA during radio/X-ray outbursts (on a scale of days) of the microquasar GRS~1915+105, which is a LMXB listed in~\citet{Liu01}.

Persistent VHE $\gamma$-ray emission from LMXBs containing a neutron star was predicted~\citep{kiraly88,cheng91}. For example, particles can be accelerated in the vicinity of accreting neutron stars, giving rise to VHE $\gamma$-ray emission through interactions of ultra-high-energy nuclei with surrounding material. No steady VHE $\gamma$-ray emission of the progenitor of GRB~060602B was obtained from our long-term data. More than a dozen LMXBs (including GRS~1915+105) and several high-mass X-ray binaries have also been observed with H.E.S.S. and no detection was seen from any of them~\citep{Dickinson08}.


\section{Conclusions}

On 2006 June 2, the first completely simultaneous observations of a
$\gamma$-ray burst (GRB~060602B) in hard X-rays and in VHE
$\gamma$-rays with an IACT instrument were obtained.

The burst position was observed with H.E.S.S. at VHE energies
before, during, and after the burst. A search for a VHE $\gamma$-ray signal coincident with the burst event, as well as before and after the burst, yielded no positive result. The 99\% confidence level flux upper limit ($>$1~TeV) for the prompt phase of GRB~060602B is $2.9\times10^{-9}\,\mathrm{erg\,cm^{-2}\,s^{-1}}$.

The nature of GRB~060602B is not yet clear, although a Galactic origin seems to be more likely. The complete and simultaneous coverage of the burst with an IACT instrument operating at VHE energies places constraints either in the Galactic X-ray binary scenario or the cosmological GRB scenario.

\acknowledgments

\footnotesize
The support of the Namibian authorities and of the University of Namibia
in facilitating the construction and operation of H.E.S.S. is gratefully
acknowledged, as is the support by the German Ministry for Education and
Research (BMBF), the Max Planck Society, the French Ministry for Research,
the CNRS-IN2P3 and the Astroparticle Interdisciplinary Programme of the
CNRS, the U.K. Science and Technology Facilities Council (STFC),
the IPNP of the Charles University, the Polish Ministry of Science and
Higher Education, the South African Department of
Science and Technology and National Research Foundation, and by the
University of Namibia. We appreciate the excellent work of the technical
support staff in Berlin, Durham, Hamburg, Heidelberg, Palaiseau, Paris,
Saclay, and in Namibia in the construction and operation of the
equipment. We acknowledge financial support by SFB~439. We thank an anonymous referee for a detailed report. This research has made use of data obtained from the High Energy Astrophysics Science Archive Research Center (HEASARC), provided by NASA's Goddard Space Flight Center. P.H. Tam thanks Keith Arnaud and Craig Gordan for assistance with the XRT analysis and acknowledges support from IMPRS-HD.


\def\arraystretch{1.3}
\tabletypesize{\normalsize}

\begin{deluxetable}{ccrcrlc}
\tablewidth{0pt}
\tablecaption{H.E.S.S. observations at the burst position \label{obspattern} }
\tablehead{ \colhead{} & \colhead{} & \colhead{} & \colhead{} & \colhead{} & \colhead{$f_\mathrm{UL}$\tablenotemark{f} } & \colhead{$f_\mathrm{UL}$\tablenotemark{f} } \\ \colhead{Date\tablenotemark{a} } & \colhead{$T_\mathrm{start}$\tablenotemark{b} } & \colhead{ Z.A.\tablenotemark{c} } & \colhead{Offset\tablenotemark{d} } & \colhead{$E_{\rm th}$\tablenotemark{e} } & \colhead{($>E_{\rm th}$)} & \colhead{($>1$~TeV)} }

\startdata

      2 & 22:03:37 & 23.3\phn  & 2.5    & 540 & \phn4.2 (7 \%) & 1.6\phn    \\
      2 & 22:33:48 & 16.5\phn  & 2.5    & 540 & 11\phd\phn (19 \%) & 4.0\phn    \\
      2 & 23:04:10 &  9.9\phn  & 2.9    & 1170 & \phn5.5 (31 \%) & 7.1\phn    \\
      2 & 23:34:10 &  3.7\phn  & 2.9    & 1060 & \phn3.3 (16 \%) & 3.6\phn    \\
      3 & 00:04:38 &  4.8\phn  & 2.1    & 240 & 20\phd\phn (11 \%) & 2.0\phn    \\
      3 & 00:34:38 & 10.6\phn  & 2.1    & 260 & \phn5.2 (3 \%) & 0.61    \\
      3 & 01:04:50 & 16.2\phn  & 1.3    & 240 & \phn8.8 (5 \%) & 0.91    \\
      3 & 01:22:02 & 22.1\phn  & 0.5    & 280 & \phn6.1 (4 \%) & 0.81    \\
      3 & 02:03:02 & 31.6\phn  & 0.5    & 320 & \phn7.4 (6 \%) & 1.2\phn    \\
      3 & 02:33:28 & 38.3\phn  & 0.5    & 460 & \phn5.8 (8 \%) & 1.7\phn    \\
      3 & 03:03:52 & 45.1\phn  & 0.5    & 600 & \phn5.5 (11 \%) & 2.4\phn    \\
      3 & 23:17:39 &  7.4\phn  & 1.0    & 220 & 11\phd\phn (5 \%) & 0.97    \\
      3 & 23:47:36 &  4.8\phn  & 1.0    & 220 & \phn4.6 (2 \%) & 0.41    \\
      4 & 00:17:46 &  8.5\phn  & 1.3    & 240 & \phn9\phd\phn (5 \%) & 0.93    \\
      4 & 00:47:46 & 14.9\phn  & 1.3    & 240 & 12\phd\phn (6 \%) & 1.2\phn    \\
      4 & 23:41:41 &  4.5\phn  & 1.2    & 220 & \phn9.3 (4 \%) & 0.83    \\
      5 & 00:12:13 &  8.9\phn  & 0.6    & 220 & \phn7\phd\phn (3 \%) & 0.60    \\
      5 & 00:42:12 & 15.1\phn  & 0.6    & 240 & \phn8.4 (4 \%) & 2.3\phn    \\
      5 & 01:12:27 & 22.9\phn  & 1.1    & 290 & 13\phd\phn (9 \%) & 1.8\phn    \\
      6 & 00:36:42 & 15.0\phn  & 0.4    & 240 & 15\phd\phn (8 \%) & 1.5\phn    \\
      6 & 01:06:48 & 21.5\phn  & 0.4    & 260 & \phn9.1 (5 \%) & 1.1\phn    \\

\enddata
  \tablenotetext{a}{ Date in 2006 June}
  \tablenotetext{b}{ Start time of the observation in UT. All but the seventh observation run, which has an exposure of 14~minutes, have an exposure time of 28~minutes.}
  \tablenotetext{c}{ Mean zenith angle of the observation run in degrees.}
  \tablenotetext{d}{ Offset of the burst position from the center of the FoV in degrees.}
  \tablenotetext{e}{ Energy threshold for a standard cut analysis in GeV.}
  \tablenotetext{f}{ 99 \% flux upper limit for a standard cut analysis in $10^{-12}$~photons~cm$^{-2}$~s$^{-1}$, assuming a photon spectral index of 2.6, where numerals in brackets indicate the fractional flux in Crab unit above the same threshold}
\end{deluxetable}

\clearpage
%

%

\end{document}